\documentclass[12pt]{article}
\catcode`\@=11
\@addtoreset{equation}{section}

\global\arraycolsep=2pt
\usepackage{mathrsfs,amsbsy,amssymb,latexsym,amsfonts,amsmath,cite}
\usepackage{graphicx,color}
\usepackage{ulem}

\newcommand{\dd}{{\rm d}}

\newcommand{\wh}{\widehat}
\newcommand{\wt}{\widetilde}

\def\half{\frac12}
\def\({\left(}
\def\){\right)}
\def\[{\left[}
\def\]{\right]}
\def \be {\begin{equation}}
\def \ee {\end{equation}}
\def \beal#1 {\begin{align}#1\end{align}}
\def \bes#1 {\begin{equation}\begin{split}#1\end{split}\end{equation}}
\def \nn {\notag\\}

\def\aver#1{\left\langle #1 \right\rangle}

\def\cD{{\cal D}}

\def\cO{{\cal O}}

\allowdisplaybreaks

\begin{document}

\begin{flushright}
\parbox{4cm}
{YITP-19-04 \\ 
KUNS-2747 \\ 
}
\end{flushright}

\vspace*{0.2cm}

\begin{center}
{\Large \bf Holographic geometry for non-relativistic systems \vspace*{0.1cm}\\ 
emerging from generalized flow equations 
}
\vspace*{1.5cm}\\
{\large  Sinya Aoki$^{\sharp}$\footnote{E-mail:~saok@yukawa.kyoto-u.ac.jp}, 
Shuichi Yokoyama$^{\sharp}$\footnote{E-mail:~shuichi.yokoyama@yukawa.kyoto-u.ac.jp}
and Kentaroh Yoshida$^{\flat}$\footnote{E-mail:~kyoshida@gauge.scphys.kyoto-u.ac.jp}} 
\end{center}

\vspace*{1cm}

\begin{center}

$^\sharp${\it Yukawa Institute for Theoretical Physics, Kyoto University, \\ 
Kitashirakawa Oiwake-cho, Sakyo-ku, Kyoto 606-8502, Japan} \\ 
$^\flat${\it Department of Physics, Kyoto University, \\ 
Kitashirakawa Oiwake-cho, Sakyo-kun, Kyoto 606-8502, Japan} 
\end{center}

\vspace{1cm}

\begin{abstract}
An intriguing result presented by two of the present authors is that an anti de Sitter space 
can be derived from a conformal field theory 
by considering a flow equation. 
A natural expectation is that given a certain data on the boundary system, 
the associated geometry would be able to emerge from a flow, {\it even beyond the conformal case}.   
As a step along this line, we examine this scenario for non-relativistic systems 
with anisotropic scaling symmetries, such as Lifshitz field theories and Schr\"{o}dinger invariant theories.  
In consequence we obtain a new hybrid geometry of Lifshitz and Schr$\ddot{\rm o}$dinger spacetimes as a general holographic geometry in this framework. We confirm that this geometry reduces to each of them by considering special non-relativistic models.
\end{abstract}

\setcounter{footnote}{0}
\setcounter{page}{0}
\thispagestyle{empty}

\newpage

\tableofcontents

\section{Introduction}

One of the fascinating subjects in String Theory is a conjectured duality  
between a string (or gravity) theory on an anti de Sitter (AdS) space in $d+1$ dimensions
and a conformal field theory (CFT) in $d$ dimensions. This is called the AdS/CFT correspondence 
\cite{M,Witten,GKP} and it is recognized as a realization of the holographic principle 
\cite{Hooft,Susskind}.  This correspondence has not been proven completely yet, 
but it is supported by a huge amount of circumstantial evidences 
and there is no contradiction so far. 

\medskip 

The validity of the AdS/CFT correspondence has been well recognized nowadays, 
and it has opened up a new arena to consider applications of AdS/CFT to realistic systems 
in condensed matter physics (CMP) (often referred to as  AdS/CMP).  
By following this approach, one can study the non-perturbative physics of gauge theories 
in a strongly-coupled region by using a weakly-coupled (semi-)classical gravity.
Indeed, a lot of works have been carried out (For nice reviews, for example, 
see \cite{Hartnoll,Herzog,McGreevy,Pires,Sachdev,Iqbal}). 
 
\medskip 
 
One of the issues in the context of AdS/CMP is to consider how to realize non-relativistic (NR) 
systems beyond the usual, relativistic CFT. A key ingredient is an anisotropic scaling like 
\begin{eqnarray}
t \rightarrow \Lambda^Z\, t, \quad x^i \rightarrow \Lambda\, x^i~~(i=1,\ldots,d-1)\,,  
\qquad \Lambda:~\mbox{a real const.}\,, 
\label{aniso}
\end{eqnarray}
where $t$ and $x^i$ are time and spatial coordinates, respectively. Here $Z$ is called 
the dynamical critical exponent, which measures anisotropy of the system. 
In particular, the $Z=1$ case corresponds to the relativistic dilatation. 
There are two famous examples of symmetry algebra including the anisotropic scaling (\ref{aniso}), 
Schr\"{o}dinger algebra \cite{Sch} and Lifshitz algebra\footnote{For more details on 
the non-relativistic algebras, see \cite{Henkel1,Henkel2,Henkel3,Henkel4,NS}. }. 
The geometries that preserve these symmetries as isometries have been constructed  
in \cite{Son,BM} and \cite{Kachru}, respectively. 
These geometries have been proposed from the symmetry argument. 
Then, with the standard dictionary (with some extension), the boundary theory is argued. 
One of the subtle points is that the boundary behavior of these geometries are not so well defined.  
Namely, a part of the geometry shrinks as we approach the boundary. Then, the supergravity 
approximation may not be valid any more and the notion of the boundary would be subtle. 

\medskip 

Based on this observation, our motivation here is to consider the inverse direction 
to the preceding works. That is, we would like to demonstrate that a gravity dual may emerge 
starting from a non-relativistic system with a scaling invariance. For this purpose, 
we shall follow the method based on flow equations proposed 
by two of the present authors\cite{Aoki:2017bru}\footnote{The flow equation method to construct a holographic theory was originally introduced in Ref.~\cite{Aoki:2015dla} and was further developed in Refs.~\cite{Aoki:2016ohw, Aoki:2016env}. }.  In fact, they have derived  an anti de Sitter space from a CFT data by considering a flow equation.  A natural expectation is that given a certain data on the boundary system, the associated geometry would be able 
to emerge from a flow, {\it even beyond the conformal case}. This possibility would open up an intriguing research arena to be studied in relation to the holographic principle. 

\medskip 

In this paper, we examine this expectation for non-relativistic systems 
with an anisotropic scaling symmetry, such as Lifshitz field theories 
and Schr\"{o}dinger invariant theories.  
As a result we obtain a hybrid geometry of Schr$\ddot{\rm o}$dinger and Lifshitz spacetimes as a general holographic space by employing the associated two-point functions as the boundary data and generalizing  the flow equation itself. This geometry contains both of them as special examples of non-relativistic models.
Our results are summarized in Table \ref{results}. 
\begin{table}[thb]
 \label{results}
 \begin{center}
  \begin{tabular}{|c|c|c|c|c|}
  \hline
 Section & Theory type &Operator type & Flow equation type & Geometry type \\
  \hline
 \ref{AdS} & CFT & CPS & R & AdS \\
 \ref{NRH} & NRCFT & NRCPS & NR & NR Hybrid \\
 \ref{NRsmearing} & CFT & CPS & NR & {Lifshitz with $Z=2$}\\
 \ref{lightconeR} & NRCFT with LS & NRCPS & R & Schr\"odinger \\
 \ref{NRdeformation} & CFT & NRCPS & R & Schr\"odinger \\
 \ref{MassEigenVector} & NRCFT & NRCPS & NR & {Lifshitz with $Z=2$}\\
 \ref{LifshitzGeometry} & Lifshitz & Lifshitz PS & Lifshitz & Lifshitz with general $Z$\\
  \hline
  \end{tabular}
 \caption{ Summary of the results in this paper. ``LS'', ``R'', ``NR'', ``(C)PS'' stand for ``light-cone reversal symmetry'', ``relativistic'', ``non-relativistic'', ``(conformal) primary scalar'', respectively.
 $Z$ denotes the dynamical critical exponent. }
 \end{center}
\end{table}

The rest of this paper is organized as follows. Section 2 provides a brief review of gravity duals for non-relativistic systems, Schr\"{o}dinger spacetimes and Lifshitz spacetimes. 
In Section \ref{AdS}, we review how an AdS geometry emerges from a CFT data given at the boundary by employing a flow equation.
In Section \ref{NRCFT}, \ref{LifshitzGeometry}, we apply the formulation to non-relativistic CFTs, a Lifshitz-type scale invariant theory, respectively.
Section \ref{discussion} is devoted to conclusion and discussion. 
The Schr\"{o}dinger algebra and the Lifshitz algebra are summarized in Appendix \ref{CSL},
while a different flow equation is considered for the Lorentzian CFT in Appendix \ref{NoncovariantFlow}. Transformation properties of the flowed field are given in Appendix \ref{formula}.

\section{Gravity duals for non-relativistic systems} 

In this section, we shall give a brief review of gravity duals for non-relativistic systems, 
Schr\"{o}dinger spacetimes and Lifshitz spacetimes. 

\subsection{Schr\"{o}dinger spacetimes}

In 2008, Son proposed a geometry \cite{Son} preserving the Schr\"{o}dinger symmetry 
as the maximal symmetry\footnote{The Schr\"{o}dinger algebra can be embedded 
into a relativistic conformal algebra as a subalgebra. Hence the usual AdS metric is 
Schr\"{o}dinger invariant, but this symmetry is not the maximal one. For the detail of the embedding, 
see Appendix A.} as a holographic dual of a non-relativistic system realized in a cold atom 
experiment. The metric is given by\footnote{The $\tau$ coordinate describes a radial direction, not the Euclidean time. } 
\begin{eqnarray}
\dd s^2 = \frac{\pm 2 \dd x^+ \dd x^- + \sum_{i=1}^{d-1} (\dd x^i)^2 + \dd \tau^2}{\tau^2} - \sigma^2 \frac{(\dd x^+)^2}{\tau^{4}}\,,  
\label{Son}
\end{eqnarray}
where $\sigma$ is a real constant parameter. This geometry can be considered
as a one-parameter deformation of a ($d+2$)-dimensional AdS space, 
which is regarded as a gravity dual of a $d$-dimensional Schr\"{o}dinger invariant system.  

\medskip 

The Schr$\ddot{\rm o}$dinger algebra \cite{Sch} is composed of a time translation, 
spatial translations, spatial rotations, Galilean boosts, a mass operator, 
a special conformal transformation and an anisotropic scaling as
\begin{eqnarray}
t \rightarrow \Lambda^{Z}\, t, \quad x^i \rightarrow \Lambda\, x^i~~(i=1,\ldots,d-1)\,,  
\qquad \Lambda:~\mbox{a real const.}\,, \label{scaling}
\end{eqnarray}
where $t$ and $x^i$ are time and spatial coordinates, respectively, 
and $Z$ is  the dynamical critical exponent. 
In the original Schr\"{o}dinger algebra, the $Z=2$ case is considered. 
In more general, one may consider an arbitrary value of $Z$\,, though 
the special conformal symmetry is broken except for $Z = 2$ or $1$. 

\medskip

It is easy to see the invariance of the metric (\ref{Son}) under the Schr$\ddot{\rm o}$dinger 
symmetry (For the detail, see \cite{Son}). In particular, the metric (\ref{Son}) is invariant 
under the scaling 
\begin{eqnarray}
x^+ \rightarrow \Lambda^2 \, x^+, \quad x^- \rightarrow x^-\,, 
\quad x^i \rightarrow \Lambda\, x^i~~(i=1,\ldots,d-1)\,.  
\quad \tau \rightarrow \Lambda \tau\, .
\end{eqnarray}

\subsection{Lifshitz spacetimes}

The Lifshitz spacetime was proposed in \cite{Kachru} as a gravity dual for the Lifshitz fixed point 
realized in condensed matter systems. (For a comprehensive review of the Lifshitz holography, see \cite{Taylor}.) The metric of the gravity dual is given by 
\begin{eqnarray}
\dd s^2 = \frac{-\dd t^{2}}{\tau^{2Z}} + \frac{\sum_{i=1}^{d-1} (\dd x^i)^2 + \dd \tau^2 }{\tau^2}\,,
\label{Lifshitz}
\end{eqnarray}
which is invariant under the rescaling 
\begin{eqnarray}
t \rightarrow \Lambda^{Z} \, t, 
\quad x^i \rightarrow \Lambda\, x^i~~(i=1,\ldots,d-1)\,,  
\quad \tau \rightarrow \Lambda \tau\,. 
\end{eqnarray}
This is a bulk realization of the scaling symmetry included in the Lifshitz algebra. 
As described in Appendix A in detail, the Lifshitz algebra is composed of a time translation, 
spatial translations, spatial rotations and the anisotropic scaling (\ref{scaling})\,.  

\medskip 

The metric (\ref{Lifshitz}) describes a $d+1$-dimensional bulk geometry 
and the boundary theory has $d-1$ spatial directions and a time direction. 

\section{Holographic geometry from a flowed conformal primary}
\label{AdS}

In Ref.~\cite{Aoki:2015dla, Aoki:2017bru}, we proposed a mechanism for a holographic geometry to emerge from a bou{n}dary QFT coarse-grained by a flow equation. 
In this section, as a warm{-}up, we demonstrate this mechanism concretely by using conformal field theories. 

\subsection{Euclidean case}
\label{EAdS}

In this subsection, we review the emergence of an AdS geometry from a generic Euclidean CFT via the flow equation approach \cite{Aoki:2017bru}. 

Consider a $D$-dimensional Euclidean CFT with a real scalar primary field $O_0(x)$ with conformal dimension $\Delta_\cO$.
By using the conformal symmetry of correlation functions, the 2-point function of the primary operator can be expressed as 
\begin{equation}
\langle  O_0(x) O_0(0) \rangle =  f_0\left( x^2 \right), \quad
f_0(\Lambda x^2) = \Lambda^{-\Delta_\cO} f_0(x^2), 
\label{f0}
\end{equation}
where $x^2:=\delta_{\mu\nu}x^\mu x^\nu$ and $f_0(u) \propto u^{-\Delta_\cO}$.

\medskip  

Let us define a flowed field $\phi_0(x;\eta)$ by a free flow equation as
\begin{equation}
\frac{ \partial \phi_0(x;\eta)}{\partial \eta} =\partial^2 \phi_0(x;\eta), \quad
 \phi_0(x; 0) =  O_0(x)\, , \qquad \partial^2 =\delta^{\mu\nu} \partial_\mu \partial_\nu ,
 \label{FlowEuclidean}
\end{equation}
where $\eta$ is a positive number called a flow parameter, and the original field is recovered at the limit $\eta\to0$.\footnote{ 
Note here that we naively take $\partial^2$ for the flow equation so as to realize a diffusion equation. This is the original choice employed in  \cite{Aoki:2015dla,Aoki:2016ohw,Aoki:2016env,Aoki:2017bru}. Only for this purpose, however, there may be some possibilities to generalize this choice, as we shall see later. 
}
The formal solution of the flow equation is given by
\begin{equation}
\phi_0(x;\eta) = e^{\eta \partial^2} O_0(x). 
\end{equation} 
This is a well-defined operator, since this can be rewritten as 
\begin{equation}
\phi_0(x;\eta) 
= \int d^D y K_0(x-y;\eta) O_0(y) \, , 
\label{flowsol}
\end{equation} 
where
\begin{equation}
K_0(x;\eta) := \frac{1}{(4\pi\eta)^{D/2}}\exp\left[- \frac{x^2}{4\eta} \right], \label{K_0ernel}
\end{equation} 
is the Green function of the flow equation:
\begin{equation}
\frac{ \partial K_0(x;\eta)}{\partial \eta} =\partial^2 K_0(x;\eta), \qquad
K_0(x;0)=\delta^{D}(x).
\end{equation} 

A virtue to coarse-grain operators by a flow equation is that the 2-point function of the flowed operator has no contact singularity.
To see this, we compute the 2-point function of the flowed field $\phi_0$ as
\begin{eqnarray}
 \langle \phi_0(x_1; \eta_1) \phi_0(x_2; \eta_2)\rangle 
 &=& {\rm e}^{\eta_1\partial^2 +\eta_2\partial'^2}
 \langle  O_0(x_1) O_0(x_2) \rangle ,
\end{eqnarray}
where we used \eqref{flowsol} and $\partial,\partial'$ act on $x_1,x_2$, respectively: $\partial_\mu := {\partial \over \partial x_1^\mu}, \partial'_\mu := {\partial \over \partial x_2^\mu}$. 
By using \eqref{f0}, we can rewrite this as 
\begin{eqnarray}
 \langle \phi_0(x_1; \eta_1) \phi_0(x_2; \eta_2)\rangle 
 &=& 
  {\rm e}^{\eta_+ \partial^2} f_0(x_{12}^2) ,
\end{eqnarray}
where $\eta_+ :=\eta_1+\eta_2$, $x_{12}:=x_1 -x_2$. 
This quantity is a smooth function of $x_{12}^2$ and $\eta_+$ since this can be rewritten by using the Green function \eqref{K_0ernel}, which implies the absence of the contact singularity, as was claimed.\footnote{ 
The explicit form was given in \cite{Aoki:2017bru}.
} 
Let us denote this smooth function by $F_{0}(x_{12}^2; \eta_+)$. Then by using the scaling relation in \eqref{f0} the function $F_{0}$ satisfies
\begin{equation}
F_{0}\left(\Lambda^2x_{12}^2; \Lambda^2 \eta_+\right)
=\Lambda^{-2\Delta_\cO} F_{0}\left(x_{12}^2; \eta_+\right). 
\end{equation}
Choosing $\Lambda=1/\eta_+^\half$ we find 
\begin{equation}
 \langle \phi_0(x_1; \eta_1) \phi_0(x_2; \eta_2)\rangle 
= \frac{1}{\eta_+^{\Delta_\cO}} F_{0}\left( \frac{x_{12}^2}{\eta_+};1\right).
\end{equation}
We introduce a normalized field denoted by $\sigma_0$ as  
\begin{eqnarray}
\sigma_0(x; \eta) := \frac{\phi_0(x; \eta) } {\sqrt{\langle \phi_0(x; \eta)^2\rangle}},
\end{eqnarray}
so that $\langle \sigma_0(x; \eta)^2\rangle =1$. 
The average is taken in the original $D$ dimensional theory. This is well-defined due to the fact that the contact singularity is resolved.
Using the normalized field we define 
\begin{eqnarray}
\hat g_{AB}(x;\eta)&\equiv&R^2 \lim_{(x',\eta')\to (x,\eta)}{\partial \over \partial z^A}\sigma_0(x; \eta)  {\partial \over \partial z'^B}\sigma_0(x'; \eta'), \\
g_{AB}(z) &\equiv& \aver{\hat g_{AB}(x;\eta)}, 
\end{eqnarray}
where $R$ is a certain length scale, 
$(z^A)=(x^\mu, \tau$) with $\tau:=\sqrt{\eta/\alpha}$ and $\alpha$ is fixed later. 
Thanks to this normalization, $g_{AB}(z)$ becomes
the information metric\cite{Aoki:2017bru}, and  thus can be regarded as a metric in the $D+1$ dimensional holographic space, called the induced metric. 
For simplicity, we take $R=1$ hereafter.

\medskip 

The 2-point function of the normalized filed $\sigma_0$ becomes 
\begin{eqnarray}
\langle \sigma_0(x_1; \eta_1)  \sigma_0(x_2; \eta_2)\rangle 
= \left(\frac{2\sqrt{\eta_1\eta_2}}{\eta_+}\right)^{\Delta_\cO} G_0\left(\frac{x_{12}^2}{\eta_+}\right),~~~~~~
\label{2ptsigma}
\end{eqnarray}
where $G_0(u) := F_0(u;1)/F_0(0;1)$, and thus $G_0(0)=1$.
Therefore the induced metric is evaluated as
\beal{
g_{\mu\nu}(z)
=&-\delta_{\mu\nu}\frac{1}{\alpha\tau^2}G_0'(0)\,, \quad 
g_{\tau\tau}(z)
=\frac{\Delta_\cO}{\tau^2}\,, 
}
and the other components vanish.

We can determine $G'(0)$ by using the flow equation for the 
2-point function of the normalized operator 
\be 
\partial_{\eta_1}\langle \sigma_0(x_1; \eta_1)  \sigma_0(x_2; \eta_2)\rangle =\partial_{x_1}^2\langle \sigma_0(x_1; \eta_1)  \sigma_0(x_2; \eta_2)\rangle, 
\ee
which leads to
\begin{eqnarray}
-\frac{\Delta_\cO}{\eta_+^{\Delta_\cO +1}} G_0\left( \frac{x_{12}^2}{\eta_+} \right) 
- \frac{x_{12}^2}{\eta_+^{\Delta_\cO +2}} G_0'\left( \frac{x_{12}^2}{\eta_+} \right) &=&
\frac{2D}{\eta_+ ^{\Delta_\cO +1}} G_0'\left( \frac{x_{12}^2}{\eta_+} \right)
+ \frac{4x_{12}^2}{\eta_+^{\Delta_\cO +2}} G_0''\left( \frac{x_{12}^2}{\eta_+} \right).~~~~~~~
\label{eq:flow_id}
\end{eqnarray}
From this we find 
\begin{equation}
G_0' (0)= -\frac{\Delta_\cO}{2D}.
\label{G'(0)}
\end{equation}
Therefore, 
taking $\alpha = 1/(2D)$, we obtain the Euclidean AdS metric as
\begin{eqnarray}
\dd s^2 &=& \Delta_\cO \frac{\dd x^2 + \dd \tau^2}{\tau^2} .
\label{eq:AdS_E}
\end{eqnarray}
Thus, the flow approach generates an AdS space from a Euclidean CFT.

This result can be confirmed from symmetry argument.
To see this let us consider the $D$ dimensional conformal transformation of the normalized field: 
\beal{
\delta^{\rm conf}\sigma_0(x;\eta)
=&- \left\{{\eta} (\partial^2 \delta x^\mu) + 2 {\eta^2}  (\partial^\nu \partial^\rho \delta x^\mu) \partial_\nu \partial_\rho   + 2{\eta} (\partial^\nu \delta x^\mu) \partial_\nu + \delta x^\mu\right\} \partial_\mu \sigma_0(x;\eta) \nn 
&- {\Delta_\cO \over D}\left\{2 {\eta} (\partial^\nu \partial_\mu {\delta x}^\mu) \partial_\nu + (\partial_\mu {\delta x}^\mu) \right\} \sigma_0(x;\eta). 
}
Following \cite{Aoki:2017bru},
we decompose this into isometries of the $D+1$ dimensional  AdS
and the rest as
\beal{
\delta^{\rm conf}\sigma_0(x;\eta)
=& \delta^{{\rm diff}} \sigma_0(x;\eta) + \delta^{{\rm extra}} \sigma_0 (x;\eta), \label{conformalTrsigma}
}
where 
\beal{
\delta^{{\rm diff}} \sigma_0(x;\eta) =& -(\bar\delta \eta \partial_\eta + \bar\delta x^\mu \partial_\mu)\sigma_0 (x;\eta), \quad 
\delta^{{\rm extra}}\sigma_0(x;\eta) 
= 4 {\eta^2} b^\nu \partial_\nu \(\partial_\eta + {\Delta_\cO +2 \over 2\eta }\)  \sigma_0 (x;\eta),
\label{extraterm}
}
with 
\be 
\bar\delta x^\mu =\delta x^\mu + 2D { \eta}  b^\mu,\quad  \bar\delta \eta =(2\lambda - 4 (b_\mu x^\mu) )\eta.
\label{AdSIsometry}
\ee
Then the conformal transformation of the induced metric is computed as 
\be
\delta^{\rm conf}  g_{AB}(z) 
=\delta^{\rm diff}  g_{AB}(z) + \lim_{(x';\eta')\to (x;\eta)} {\partial \over \partial z^A} {\partial \over \partial z'^B}\aver{ \delta^{{\rm extra}}\sigma_0(x;\eta) \sigma_0(x';\eta')  +\sigma_0(x;\eta) \delta^{{\rm extra}}\sigma_0(x';\eta') }. 
\label{eq:confTrMetric}
\ee
where 
\be
\delta^{\rm conf}  g_{AB}(z) := \aver{\delta^{\rm conf} \hat g_{AB}(x;\eta) }, \quad 
\delta^{\rm diff}  g_{AB}(z) := \aver{\delta^{\rm diff} \hat g_{AB}(x;\eta) }.
\ee
Since the 2-point correlation function is invariant under an arbitrary conformal transformation, $\delta^{\rm conf}  g_{AB}(z) = 0$. On the other hand, by using \eqref{2ptsigma} and \eqref{extraterm} we find 
\begin{eqnarray}
&&\aver{\delta^{{\rm extra}}\sigma_0(x_1;\eta_1) \sigma_0(x_2;\eta_2)  + \sigma_0(x;\eta)\delta^{{\rm extra}}\sigma_0(x_2;\eta_2)  } \nonumber \\
&=& -8  {(\sqrt{4 \eta_1 \eta_2})^{\Delta_\cO} \over  \eta_+^{\Delta_\cO+2}}(\eta_1 -\eta_2) (b \cdot x_{12}){x_{12}^2}
 G_0''\left({x_{12}^2 \over { \eta_+}}\right),  
\end{eqnarray}
where we set $\alpha\cdot\beta := \alpha^\mu\beta_\mu$.
This implies that the 2nd term in \eqref{eq:confTrMetric} vanishes.  
Therefore 
\be 
\delta^{\rm diff}  g_{AB}(z) = 0. 
\ee
Since this result means that the metric is invariant under the isometry of AdS ($\delta^{\rm diff}$, $\bar\delta$), 
which is a maximally symmetric space, 
the induced metric has to be an AdS one up to an overall constant.
This is the desired result.

\subsection{Lorentzian case}
\label{LAdS}

In this subsection we comment on the construction of a Lorentzian AdS geometry by a flow equation approach. 
To this end the first thing to do is to smear operators by a flow equation in the Lorentzian space. This should be done so as to preserve the causal structure at least in a certain amount. 
Although it is unclear whether such a smearing is possible by a diffusion-type differential equation, a natural candidate of a Lorentzian free flow equation may be given by 
\begin{equation}
\frac{ \partial \phi_0(x;\eta)}{\partial \eta} =\partial^2 \phi_0(x;\eta) , \qquad \partial^2 = g^{\mu\nu}\partial_\mu \partial_\nu , 
\label{FlowLorentzian}
\end{equation}
with $g^{\mu\nu}={\rm diag}(-1, 1,\cdots, 1)$. 
However it is soon realized that the formal solution of this flow equation becomes divergent. 

To avoid this problem, in what follows, we simply use the Wick rotation for a time coordinate. Then the Lorentzian flow equation \eqref{FlowLorentzian} is mapped in the Euclidean one \eqref{FlowEuclidean}, which can be solved without any problems.\footnote{The AdS/CFT correspondence in the Lorentzian space has a different aspect from the Euclidean case \cite{Balasubramanian:1998sn,Banks:1998dd,Balasubramanian:1998de}.
For example, see \cite{Marolf:2004fy} for a careful study on the analytic continuation in this context. 
} Then we rotate the result back to the Lorentzian space.%
\footnote{ 
This approach has a virtue that Lorentz invariance is manifest, but assumes a sufficient fall-off of a flowed operator at infinity, which is non-trivial for any operator in CFT. We instead present a different approach to this problem in appendix \ref{NoncovariantFlow}. }

Let us start the flow equation in the Euclidean space \eqref{FlowEuclidean}. 
For later purpose we introduce a complex coordinate that $x=(\vec x, x^+, x^-)$ with $x^{\pm} = (x^{D-1}\pm ix^D)/\sqrt{2}$
and rewrite the flow equation as
\begin{equation}
\frac{ \partial \phi_0(x;\eta)}{\partial \eta} =\partial^2 \phi_0(x;\eta) , \qquad \partial^2= \vec \partial^2 + 2\partial_+\partial_-. 
\end{equation}
The flowed field is expressed as
\begin{equation}
\phi_0(x;\eta) = \int d^D y K_0(x-y;\eta) O_0(y), \quad
K_0(x;\eta) = \frac{1}{(4\pi\eta)^{D/2}}\exp\left[- \frac{\vec x^2 + 2x^+x^-}{4\eta} \right] .
\end{equation} 

The 2-point function of the normalized field is given by 
\begin{eqnarray}
\langle \sigma_0(t_1,\vec x_1; \eta_1) 
 \sigma_0(t_2,\vec x_2; \eta_2)\rangle 
 = \left(\frac{2\sqrt{\eta_1\eta_2}}{\eta_+}\right)^{\Delta_\cO} G_0\left(\frac{\vec x^2 + 2x^+x^-}{\eta_+}\right),~
\end{eqnarray}
which leads to
\begin{eqnarray}
g_{ij} &=& -\delta_{ij}\frac{G_0'(0)}{\alpha\tau^2}, \quad
g_{+-}=g_{-+}=-\frac{G_0'(0)}{\alpha\tau^2}, \quad
g_{\tau\tau} = \frac{\Delta_\cO}{\tau^2}, 
\end{eqnarray}
and the others vanish.  
By using \eqref{G'(0)} and $\alpha = 1/(2D)$, this gives the Euclidean AdS metric containing a complex coordinate
\begin{eqnarray}
\dd s^2 &=& \Delta_\cO \frac{\dd \vec x^2 + 2\dd x^+ \dd x^- + \dd\tau^2}{\tau^2}.
\end{eqnarray}
By the analytic continuation $x^D= - ix^0$, the coordinates $x^+, x^-$ become the light-cone ones $x^\pm =(x^{D-1} \pm x^0)/\sqrt{2}$, which converts the Euclidean AdS metric to the Lorentzian one with the  formally same expression.

\section{Holographic geometry from flowed non-relativistic conformal primaries}
\label{NRCFT}
In this section we shall apply the flow equation approach (presented in section \ref{AdS}) to a non-relativistic conformal primary operator, and investigate induced geometry by a non-relativistic flow equation.

\subsection{General induced geometry}
\label{NRH}

In this subsection we investigate a general holographic geometry for a generic non-relativistic CFT with a non-relativistic flow equation. To this end we start with a $d$-dimensional non-relativistic CFT with a primary scalar field $O(\vec x ,t)$ with a general dimension $\Delta_\cO$. 
The non-relativistic conformal symmetry constrains the 2-point function of this operator as 
\beal{
\aver{O(\vec x_1,t_1)O^\dagger(\vec x_2,t_2)} ={1\over t_{12} ^{\Delta_\cO}}f({\vec x_{12}^2 \over 2t_{12} } ), 
\label{2ptNRCFT}
}
where $\vec x^2=x_ix^i$. 

For our argument,  let us introduce an extra direction denoted by $x^-$, 
in order to embed the $d$-dimensional non-relativistic symmetry into
the $D =d+1$ dimensional relativistic conformal symmetry, generated with
\be
\delta x^\mu = a^\mu + \omega^\mu{}_\nu x^\nu + \lambda x^\mu + b^\mu x^2 -2 x^\mu (b \cdot x),
\label{conftr}
\ee
where $a^\mu$, $\omega^{\mu\nu}$, $\lambda$ and $b^\mu$ are parameters of translation, rotation, dilatation and special conformal transformation, respectively, and the $D$ dimensional light-cone coordinate is given by  $ x:= (x^\mu)= (\vec x, x^+, x^-)$ with $x^+=t$. 
The transformation law under the Schr\"odinger symmetry can be derived 
from the conformal transformation (\ref{conftr}) for 
a scalar primary operator with dimension $\Delta_\cO$,
\begin{eqnarray}
\delta^{\rm conf} O_0(x) &=& -\delta x^\mu \partial_\mu O_0(x) -\frac{\Delta_\cO}{D} (\partial_\mu  \delta x^\mu)\, O_0(x)\,,
\label{eq:primary}
\end{eqnarray}
by keeping only the following components of 
the parameters,
\be 
a^i\,,  \quad a^+\,,  \quad a^- \equiv \mu, \quad \omega^i{}_j\,,  \quad  
\omega^{-i} \equiv v^i\,,  \quad  \omega^{+-} \equiv \lambda\,, \quad b^- \equiv b\,,  
\label{para-Sch}
\ee
and the other parameters are set to zero. 
Then the transformation law \eqref{conftr} reduces to 
\bes{ 
\delta^{s} x^i =& a^i + \omega^{i}{}_j x^j  - v^i x^+  + \lambda x^i -2 b x^i  x^+\,, \\
\delta^{s} x^+ =& a^+ + 2\lambda x^+ - 2b (x^+)^2\,, \\
\delta^{s} x^- =& \mu + v^i x^i  + b \vec x^2 \,. 
\label{SchrodingerTr}
}
Here $v^i$, $\mu$\,, $\lambda$ and $b$ correspond to parameters of the Galilei transformation, 
mass operator, scale transformation and special conformal transformation, respectively. 
Note here that the $x^-$ coordinate does not appear on the right-hand sides. This means that the 
translation for the $x^-$ direction commutes with the other transformations, 
so that the associated generator $i\partial_-$ is identified with the mass operator. 
Notably, $x^+$ corresponds to the time direction in the  non-relativistic system.

\medskip 

Thus a scalar primary operator $O(x)$ with dimension $\Delta_\cO$ transforms as 
\beal{
\delta^{S} O(x) =&\big[- a^i \partial_i- a^+\partial_{+}-{\mu}\partial_- -\omega^{ij} x^j\partial_i -v^i(x_i \partial_- -x^+ \partial_{i})-\lambda (\Delta_{O} + x^i \partial_i +2x^{+} \partial_{+} ) \nn
&+ b\{ - (x^i)^2 \partial_- +2 x^+(\Delta_\cO + x^i \partial_i+x^{+} \partial_{+})  \} \big] O(x)\,,
\label{SchrodingerPrimary}
}
and its 2-point function is given by
\beal{
\aver{O(\vec x_1,x_1^+,x_1^{-})O^\dagger(\vec x_2,x_2^+,x_2^{-})} ={1\over (x_{12}^+)^{\Delta_\cO}}f\(x_{12}^-+{\vec x_{12}^2 \over 2x_{12}^+ } \). 
\label{2ptNRCFTx-}
}
It is easy to see that \eqref{2ptNRCFTx-} reduces to \eqref{2ptNRCFT} when $x_{12}^-\to0$.

We smear this non-relativistic conformal primary operator by a non-relativistic flow equation. 
A general non-relativistic free flow equation is 
\be 
{\partial \over \partial \eta}\phi(x;\eta) =(2i\bar m \partial_+ + 2\partial_- \partial_+ + \vec\partial^2) \phi(x;\eta) , \quad \phi(x;0) = O(x), 
\label{NRflow}
\ee
where $\bar m$ is a real parameter of mass dimension one.  
This can be solved as 
\beal{  
\phi(x;\eta) 
=& e^{\eta  (2i\bar m\partial_+ + 2\partial_- \partial_+ + \vec\partial^2)} O(x) 
= \int d^Dx^{\prime} K(x-x';\eta) O(x')
}
where 
\beal{
K(x;\eta) 
=\exp\[{\displaystyle -i\bar m x^-  }\] {e^{-2  x^+ x^- - \vec x^2 \over 4\eta  }\over \sqrt{4\pi \eta}^{D}}.
}

Let us study the transformation of the flowed operator under the non-relativistic conformal transformation.  
\beal{
\delta_{\bar m}^S \phi(x;\eta) 
=  e^{\eta  (2i\bar m \partial_+ + 2\partial_- \partial_+ + \vec\partial^2)} \delta^S O(x)
= \delta^S \phi(x;\eta) + \delta_{\bar m}'^S \phi(x;\eta)
\label{deltaSM}
}
where $\delta^S$ is given by \eqref{SchrodingerPrimary} and 
\beal{
\delta_{\bar m}'^S \phi(x;\eta)
=&\big[ 2i  \bar m \eta v^i\partial_i   -\lambda 2\eta \partial_\eta + b\{(- 2\eta(d-1) +4\eta \Delta_\cO +4\eta^2\partial_\eta)\partial_- \nn
&+4i\bar m \eta (\Delta_\cO +   x^i\partial_i + \eta\vec\partial^2 +\eta \partial_\eta) +4x^+\eta\partial_\eta  
 \} \big] \phi(x;\eta) .
}
Note that the flowed operator $\phi(x;\eta)$ transforms differently from $O(x)$ under the Galilei transformation unless $\bar m$ vanishes. 

We move on to the 2-point function of the flowed operator, which is written as 
\beal{ 
\aver{\phi(x_1;\eta_1)\phi^\dagger(x_2;\eta_2)} 
=& e^{\eta_1(2i\bar m \partial_+ + 2\partial_- \partial_+ + \vec\partial^2) +\eta_2(-2i\bar m \partial'_+ + 2\partial_-' \partial_+' + \vec\partial'^2)}\aver{O(x_1)O^\dagger(x_2)} \nn
=& \exp\[{\displaystyle\eta_+(2i\bar m \partial_+ + 2\partial_- \partial_+ + \vec\partial^2)}\]\bigg[ {f\(x_{12}^{-}+{ \vec x_{12}^2\over 2x^+_{12}}\) \over (x^+_{12} )^{\Delta_\cO}}\bigg]
}
where we used the same notation in the previous section.
This function can be written as $F_1( (\vec x_{12})^2,x_{12}^+, x_{12}^{-};\eta_+)$. From the Galilean invariance we find 
\beal{
v^i ( x^i \partial_{-} -  (x^++ 2i\bar m \eta){2 x^i }\partial_{\vec x^2}) F_1( \vec x^2,x^+, x^{-};\eta) = 0. 
}
This can be generally solved by 
\beal{
F_1( \vec x^2,x^+, x^{-};\eta)= F(2(x^+ + 2i\bar m \eta)x^{-}+ \vec x^2,x^+;\eta ) ,
}
where $F$ is an unknown function. 
The non-relativistic scale invariance requires  
\be 
 F_1(\lambda^2 \vec x^2, \lambda^2 x^+, x^{-};\lambda^2\eta_+)= \lambda^{-2\Delta_\cO} F_1(\vec x^2, x^+, x^{-};\eta_+) ,
\ee
which constraints the function in such a way that  
\be 
F(2( x^++ 2i\bar m \eta_+) x^{-}+ \vec x^2 , x^+;\eta_+ ) = {1\over \eta_+^{\Delta_\cO}} F\({2(  x^+ + 2i\bar m \eta_+) x^{-}+ \vec x^2\over \eta_+} ,  {  x^+ \over \eta_+} ;1 \) .
\ee
As a result the 2-point function is written as 
\be 
\aver{\phi(x_1;\eta_1)\phi^\dagger(x_2;\eta_2)} 
= {1\over \eta_+^{\Delta_\cO}} F\({2( x_{12}^+ + 2i\bar m \eta_+) x_{12}^{-}+ (\vec x_{12})^2\over \eta_+} ,  { x_{12}^+ \over \eta_+} ;1\) .
\label{2ptNRflowedOp}
\ee
Note that the contact singularity is resolved with a general parameterization.  

By using this result the 2-point function of the normalized field becomes
\beal{
\aver{\sigma(x_1;\eta_1) \sigma^\dagger(x_2;\eta_2) } 
=& \left({4\eta_1\eta_2 \over \eta_+^2}\right)^{\Delta_\cO/2} G\left({2( x_{12}^+ + 2i\bar m \eta_+) x_{12}^{-}+ (\vec x_{12})^2\over \eta_+} ,  { x_{12}^+ \over \eta_+}\right) ,
\label{2ptgeneralFlowEq}
}
where $G(u,v) =F(u,v;1)/F(0,0;1) $. 

{Since the normalized flowed field $\sigma$ is complex-valued for non-relativistic theories in general, the definition of the metric operator, for example, should be modified like 
\be
\hat g_{AB} (x;\eta) = { \partial_A \sigma(x;\eta) \partial_B \sigma^\dagger(x;\eta) + \partial_B \sigma(x;\eta) \partial_A \sigma^\dagger(x;\eta) \over 2 }, 
\label{DefinitionMetricOperator}
\ee
so that the induced metric, which is given by $ g_{AB} (z) = \aver{\hat g_{AB} (x;\eta)}$\,, becomes real and symmetric.\footnote{{ 
If one wants to relate the induced metric to an information metric for a complex-valued vector model as in \cite{Aoki:2017bru}, an extra term $\frac{1}{2}( \aver{\sigma^\dagger\partial_A \sigma}\aver{\sigma^\dagger\partial_B \sigma}+ \aver{\partial_A \sigma^\dagger\sigma}\aver{\partial_B \sigma^\dagger\sigma})$ 
is necessary to add to the definition of the induced metric. However, this term may not be written as an expectation value of a specific operator. Hence we shall avoid using this definition here, probably though the relevance to an information metric would be significant for the bulk description as shown in \cite{Aoki:2017bru}. We shall leave this issue as a future work. We appreciate Janos Balog for discussion on this point. }
}
} 
Then the induced metric is computed as 
\beal{ 
g_{\eta\eta}(z) 
=&{\Delta_\cO \over 4\eta^2 } , \quad 
g_{+\eta}(z) \,
 =  \, 
g_{-\eta}(z) 
=0,\\
g_{++}(z) 
=&{-1 \over 4\eta^2} G^{(0,2)}(\vec0), \quad 
g_{+-}(z) 
={-G^{(1,0)}(\vec0) -2i\bar m G^{(1,1)}(\vec0)\over \eta} , \quad \nn
g_{--}(z) 
=&-(4i\bar m)^2 G^{(2,0)}(\vec0), \quad 
g_{ij}(z) 
={-\delta_{ij} \over \eta}G^{(1,0)}(\vec0),
}
where $G^{(n,m)}(u,v) := \partial^n_u \partial^m_v G(u,v)$.
{The undetermined constants are not independent from each other}, since 
the flow equation implies
\be
\partial_\eta \aver{\phi(x;\eta)\phi^\dagger(x';\eta')}= (2(i\bar m + \partial_-) \partial_{+} + \partial_i^2)\aver{\phi(x;\eta)\phi^\dagger(x';\eta')}. 
\label{2ptFlowEq}
\ee
From this we obtain 
\begin{eqnarray}
 -\Delta_\cO &=& (2d +2)G^{(1,0)}(\vec0)  + 8i\bar m G^{(1,1)}(\vec0)+ 2i\bar m G^{(0,1)}(\vec0) ,\label{2ptFlowConstraintA}
\\
(-\Delta_\cO -1) G^{(1,0)} (\vec0) 
&=& {(2d + 6 )} G^{(2,0)}(\vec0) + 8i\bar m G^{(2,1)}(\vec0)+ 2i\bar m G^{(1,1)}(\vec0).
\label{2ptFlowConstraintB}
\end{eqnarray}

The induced line element is thus written as
\beal{ 
ds^2 =& {\Delta_\cO \over 4\eta^2} d\eta^2  + {-G^{(0,2)}(\vec0) \over 4\eta^2}(dx^+)^2 + 2 {-G^{(1,0)}(\vec0) -2i\bar m G^{(1,1)}(\vec0) \over \eta} dx^+dx^- \nn
&+(4\bar m)^2G^{(2,0)}(\vec0) (dx^-)^2 +{-\delta_{ij} G^{(1,0)}(\vec0) \over \eta} dx^idx^j. 
\label{generalNRgeometry}
}
We refer to this geometry as the Non-relativistic (NR) Hybrid geometry.
In what follows, we apply this result to specific examples.

\subsection{Non-relativistic flow of a conformal primary}
\label{NRsmearing}

A first application is to non-relativistic smearing of a conformal primary scalar operator in a general CFT considered in Section \ref{AdS}. 
In this case the induced geometry is more simplified than  \eqref{generalNRgeometry}.

To see this let us smear the conformal primary scalar field $O_0(x)$ by the non-relativistic flow equation \eqref{NRflow} and denote the smeared operator by $\phi_{\bar m}(x;\eta)$. 
The relation between the relativistic flowed operator and the non-relativistic one is 
\be 
\phi_{\bar m}(x;\eta) =e^{2i\eta\bar m\partial_+} \phi_0(x;\eta).
\label{eq:flow_m}
\ee
Therefore the 2-point function of the flowed operator is 
\beal{
\langle \phi_{\bar m}(x_1;\eta_1) \phi_{\bar m}^\dagger(x_2;\eta_2) \rangle
=&e^{2i\eta_1\bar m\partial_+} e^{-2i\eta_2\bar m\partial'_+} \langle \phi_0(x_1;\eta_1) \phi_0(x_2;\eta_2) \rangle \nn
=& {1\over \eta_+^{\Delta_\cO}} F_0\({2(x_{12}^+ + 2i\eta_+\bar m)x_{12}^- +(\vec x_{12})^2\over \eta_+} ;1 \) , 
}
and that of the normalized field is 
\begin{eqnarray}
\langle \sigma_{\bar m}(x_1;\eta_1) \sigma_{\bar m}^{{\dagger}}(x_2;\eta_2) \rangle &=&\left( \frac{2\sqrt{\eta_1\eta_2}}{\eta_+}\right)^{\Delta_\cO} G_0\left({2(x_{12}^+ + 2i\eta_+\bar m)x_{12}^- +(\vec x_{12})^2\over \eta_+} \right). 
\end{eqnarray}
$G(u,v)$ in Section \ref{NRH} now reduces to $G_0(u)$ so that
\be
G^{(n,0)}(\vec 0) = G_0^{(n)}(0), ~~ G^{(n,k)}(\vec 0) = 0, 
\ee
for $n\in{\mathbb Z}_{\geq0} , k\in{\mathbb Z}_{\geq1}$. 
Then \eqref{2ptFlowConstraintB} reduces to 
\be
G^{(2,0)}(\vec0) = -\frac{\Delta_\cO +1}{2(D+2)} G'(0)
= \frac{\Delta_\cO(\Delta_\cO+1)}{4D(D+2)}
\ee 
where $D=d+1$. As a result the induced line element given by \eqref{generalNRgeometry} reduces to 
\beal{ 
ds^2 =& \Delta_\cO \bigg[{\gamma} (dx^-)^2 + { d\tau^2 + 2 dx^+dx^-  + d\vec x^2 \over \tau^2} \bigg] 
\label{4.32}
}
{where we set
\begin{equation}
\tau \equiv \sqrt{2D \eta}\,, \qquad \gamma \equiv 4\bar m^2 \frac{\Delta_\cO +1}{D(D +2)}\,, 
\end{equation}
both of which are positive.} 
Note that the limit $\bar m\to0$ reduces this geometry to the AdS one given in the previous section. 

{The metric (\ref{4.32}) is a Schr$\ddot{\rm o}$dinger spacetime with $Z=0$ and the wrong sign.  
It is well known that this geometry can be regarded as a Lifshitz geometry with $Z=2$ after an appropriate compactification \cite{Donos:2010tu,Horowitz:2011gh}. This can be easily seen by performing the completing square with respect to $dx^-$. The resulting metric is given by 
\be 
ds^2 = \Delta_\cO \bigg[- {(dx^+)^2 \over \gamma\tau^4}+ { d\tau^2  + d\vec x^2 \over \tau^2} +\gamma (dx^- + {dx^+ \over \gamma\tau^2} )^2 \bigg]. 
\ee
By taking a compactification along the $x^-$-direction, this metric describes a Lifshitz spacetime \cite{Donos:2010tu,Horowitz:2011gh}. 
}

\subsection{Light-cone reversal symmetry with the relativistic flow }
\label{lightconeR}

As another simple example, we consider a holographic geometry emerging from the non-relativistic theory with the light-cone reversal symmetry ($x^\pm\rightarrow -x^\pm$) generated by the relativistic flow ($\bar m=0$). 
This implies that the 2-point function of the normalized field given by \eqref{2ptgeneralFlowEq} is invariant under $x_{12}^\pm \rightarrow -  x_{12}^\pm$,  so that we can write $G(u,v) = \wh G(u,v^2)$ with an unknown function $\wh G$.
Therefore, we have
\begin{eqnarray}
G^{(n,1)}(\vec 0) &=& 0, \quad G^{(0,2)}(\vec 0) = 2 \wh G^{(0,1)}(\vec 0),\quad
G^{(1,0)}(\vec 0) = -\frac{\Delta_\cO}{2 D} .
\end{eqnarray}
Setting $\eta =\dfrac{\tau^2}{2D}$, we obtain
\begin{eqnarray}
ds^2 &=& \Delta_\cO \[ - \frac{2 D^2 \wh G^{(0,1)}(\vec 0)}{\Delta_\cO}\frac{(dx^+)^2}{\tau^4}
+\frac{d\tau^2 + 2dx^+ dx^- + d\vec x^2}{\tau^2}
\label{MetriclightconeR}
\].
\end{eqnarray}
This metric describes nothing but a Schr\"odinger spacetime with $\sigma^2 :=  \dfrac{2 D^2\wh G^{(0,1)}(\vec 0)}{\Delta_\cO}$ studied in Ref.~\cite{Son}. 

\subsection{Non-relativistic deformation of a conformal primary}
\label{NRdeformation}
As a more non-trivial example, we consider a non-relativistic deformation of a conformal primary scalar operator in a relativistic CFT which preserve the property of the non-relativistic conformal primary condition.

Let us consider a general CFT in Section \ref{AdS} and deform the conformal primary scalar field $O_0(x)$ as
\begin{equation}
O_\epsilon(x) := e^{\epsilon \partial_-^2} O_0(x),
\label{NRDeformation}
\end{equation}
where $\epsilon$ is a real deformation parameter. 
{In order for the deformed operator to be well-behaved, the parameter $\epsilon$ needs to be positive. This will turn out to be important to obtain a Schr\"odinger space-time with the correct sign.}

Let us show that the deformed operator $O_\epsilon(x)$ is a conformal primary operator in the non-relativistic conformal algebra. 
The conformal transformation of $O_\epsilon(x)$ is given by
\begin{eqnarray}
\delta_\epsilon^{\rm conf} O_\epsilon(x) &=& e^{\epsilon\partial_-^2} \delta^{\rm conf} O_0(x) 
= \delta^{\rm conf} O_\epsilon (x) + \delta_\epsilon' O_\epsilon (x),
\end{eqnarray}
where $\delta^{\rm conf}$ is given by \eqref{eq:primary} with \eqref{conftr}, and 
\begin{align}
\delta_\epsilon' O_\epsilon (x) =&2\epsilon \big[
(2b \cdot x -\lambda -\omega^{-+}) \partial_-^2  -\omega^{i+}\partial_i\partial_- \nn
&+2(x \cdot \partial +\Delta_\cO +1) b^+\partial_- -2x^+b\cdot\partial \partial_-  +  4 \epsilon b^+ \partial_-^3 \big] O_\epsilon (x) .
\end{align}
$\delta_\epsilon' O_\epsilon$ 
vanishes if and only if 
\be 
\omega^{+-}=\lambda, ~~ \omega^{i+}= b^i = b^+ =0.
\label{commutingcase}
\ee
The subalgebra with this parameter constraint is nothing but the Schr\"odinger algebra given in \eqref{SchrodingerTr}.
In other words, the subalgebra commuting with the operator $e^{\epsilon\partial_-^2}$ becomes the Schr\"odinger algebra. 
Therefore the deformed operator $O_\epsilon(x)$ is a non-relativistic conformal primary operator. 

Hence we can apply the result in Section~\ref{NRH} to this deformed operator. 
The induced geometry obtained from the flowed operator of $O_\epsilon$ by the non-relativistic flow equation \eqref{NRflow} is given by \eqref{generalNRgeometry}, where the metric components are determined as a function of $\epsilon$.

An interesting situation happens when $\bar m=0$, where the flow equation becomes relativistic. In this case both CFT and the flow equation enjoy the light-cone reversal symmetry, so that the situation reduces to Section \ref{lightconeR}, and the induced metric reduces to a Schr\"odinger one given by \eqref{MetriclightconeR}:
\be
ds^2 = \Delta_\cO \[ - \frac{2 D^2 \wh G_\epsilon^{(0,1)}(\vec 0)}{\Delta_\cO}\frac{(dx^+)^2}{\tau^4}
+\frac{d\tau^2 + 2dx^+ dx^- + d\vec x^2}{\tau^2}
\] ,
\ee
where $\wh G_\epsilon^{(0,1)}$ describes the 2-point function of the normalized flowed field:
\be 
\langle \sigma_\epsilon(x_1;\eta_1) \sigma_\epsilon(x_2;\eta_2) \rangle
= \left( \frac{2\sqrt{\eta_1\eta_2}}{\eta_+}\right)^{\Delta_\cO}  \wh G_\epsilon\({2 x_{12}^+ x_{12}^{-}+ (\vec x_{12})^2\over \eta_+} , \({ x_{12}^+ \over \eta_+}\)^2  \) .
\ee

In the current case, we can evaluate $\wh G_\epsilon^{(0,1)}(\vec 0)$. 
To this end let us investigate the transformation rule of the flowed field $\phi_\epsilon(x;\eta)$ under the conformal transformation:
\begin{eqnarray}
{\delta}_\epsilon^{\rm conf}\phi_\epsilon(x;\eta) &=& e^{\eta(\vec \partial^2 + 2\partial_+\partial_-)+\epsilon\partial_-^2}\, \delta^{\rm conf}O(x) = \delta^{\rm conf}\phi_\epsilon(x;\eta) +\delta_\epsilon' \phi_\epsilon(x;\eta) 
\end{eqnarray}
where 
\begin{eqnarray}
\delta_\epsilon'  \phi_\epsilon(x;\eta)&:=& \big[
2\eta \{ \left( 2 b\cdot x -\lambda\right) \partial_\eta 
-(d-1-2\Delta_\cO) b\cdot \partial \} \nn
&&+ 2\epsilon \left\{ (2 b\cdot x -\lambda-\omega^{-+}) \partial_-^2 +2(x\cdot\partial +\Delta_\cO+1) b^+\partial_- -\omega^{i+}\partial_i\partial_-  -2 x^+\partial_- b\cdot \partial \right\} \nn
&&+ 4\eta^2  b\cdot \partial \partial_\eta+ 8 \epsilon\eta b^+\partial_- \partial_\eta + 8\epsilon^2 b^+\partial_-^3 \big] \phi_\epsilon(x;\eta).
\end{eqnarray}
If we restrict the conformal transformation to the Schr\"odinger one, \eqref{commutingcase}, then the terms dependent on $\epsilon$ drop out and we have \eqref{deltaSM} with $\bar m=0$.
Since the 2-point function is invariant under the rotation parametrized by $\omega^{i+}$, which is outside the Schr\"odinger algebra, we find
\begin{eqnarray}
\left(x^{-}\partial_i -x^i\partial_+ + 4\epsilon \partial_- \partial_i\right) \wh G_\epsilon\left(\frac{\vec x^2+2x^+ x^- }{\eta}, \frac{(x^+)^2}{\eta^2} \right) &=& 0, 
\end{eqnarray}
which leads to 
 \begin{eqnarray}
\wh G_\epsilon^{(0,1)}(u, v) &=& 8\epsilon \wh G_\epsilon^{(2,0)}(u, v).
\end{eqnarray}
On the other hand, \eqref{2ptFlowConstraintB} gives 
\begin{eqnarray}
\wh G_\epsilon^{(2,0)}(\vec0) &=& -\frac{\Delta_\cO +1}{2(D+2)} \wh G_\epsilon^{(1,0)}(\vec 0)
= \frac{\Delta_\cO(\Delta_\cO+1)}{4D(D+2)}.
\end{eqnarray}
Therefore we obtain 
\begin{eqnarray}
\wh G_\epsilon^{(0,1)}(\vec0) &=& 2\epsilon  \frac{\Delta_\cO(\Delta_\cO+1)}{D(D+2)}. 
\label{f01F20}
\end{eqnarray}
Finally the induced line element is obtained as 
\be
ds^2 = \Delta_\cO \bigg[-\epsilon{4(\Delta_\cO +1) D \over D+2} {(dx^+)^2\over \tau^4} + { d\tau^2 + 2 dx^+dx^-  + d\vec x^2 \over \tau^2} \bigg].
\ee
{Since the parameter $\epsilon$ is positive, this is a Schr\"odinger space-time with the correct sign.}

This result is in fact guaranteed by the symmetry. 
Let us restrict the argument of conformal symmetry in the previous section to that of the non-relativistic one. Then the normalized field $\sigma_\epsilon(x;\eta)$ transforms under the Schr\"odinger transformation as
\begin{eqnarray}
\delta_\epsilon^S \sigma_\epsilon(x;\eta) &=& \delta^{\rm diff}_S\sigma_\epsilon(x;\eta) + \delta^{\rm extra} \sigma_\epsilon(x;\eta), 
\end{eqnarray}
where $ \delta^{\rm diff}_S$ generates isometries of the Schr\"odinger spacetime as
\begin{eqnarray}
 \delta^{\rm diff}_s\sigma_\epsilon(x;\eta) &=&- \bar\delta^s\! x^A \partial_A \sigma_\epsilon(x;\eta), \\
 \bar\delta^{s} x^i &=& a^i +\omega^{i}{}_j x^j - v^i x^+ +\lambda x^i -2b x^+ x^i, \quad
\bar\delta^{s} x^+ = a^+ + 2\lambda x^+ - 2b (x^+)^2, \nonumber\\
\bar\delta^{s} x^- &= & \mu + v^i x^i + b (\vec x^2 + \tau^2), \quad 
\bar\delta^{s} \tau = (2b x^+ -\lambda) \tau, \nonumber
\end{eqnarray}
while the extra contribution becomes
\begin{eqnarray}
\delta^{\rm extra} \sigma_\epsilon(x;\eta) &=& 4b \eta^2 \partial_-\(\partial_\eta +\frac{\Delta_\cO+2}{2\eta}\)  \sigma_\epsilon(x;\eta). 
\end{eqnarray}
From this it follows that 
\begin{eqnarray}
\delta^{\rm extra} \langle \sigma_\epsilon(x_1;\eta_1) \sigma_\epsilon(x_2;\eta_2) \rangle
&=& -8b x^+_{12} x_{12}^2 \frac{(\eta_1-\eta_2)}{\eta_+}\(\frac{2\sqrt{\eta_1\eta_2}}{\eta_+}\)^{\Delta_\cO} \wh G_\epsilon^{(2,0)}\(\frac{x_{12}^2}{\eta_+}, \frac{(x_{12}^+)^2}{\eta_+^2}\) .
~~~~~~~
\end{eqnarray}
Therefore, we have $\delta^{\rm extra} g_{AB}(z) =0$, which implies $\delta^{\rm diff}_S g_{AB}(z) =0$. This shows that the resulting induced metric is invariant under the transformations forming the Schr\"odinger algebra, which requires the geometry to be a Schr\"odinger spacetime.

\subsection{Mass eigenvector}
\label{MassEigenVector}

Finally we consider the case where a conformal primary scalar in a non-relativistic CFT becomes a mass eigenvector such that
\be 
[M, O_m] = m O_m, \quad M= i\partial_-
\ee
with a mass parameter $m$, and the transformation of the operator $O_m$ is given by \eqref{SchrodingerPrimary}. 
Then the 2-point function of this primary operator given by  \eqref{2ptNRCFTx-} is now more constrained as \cite{Henkel1,NS}
\be 
\aver{O_m(\vec x_1,x_1^+,x_1^{-})O_m(\vec x_2,x_2^+,x_2^{-})^\dagger} = {C \exp\[ im \(x_{12}^{-}+{\vec x_{12}^2\over 2x_{12}^+}\)\] \over (x_{12}^+)^{\Delta_\cO}}
\label{2ptNRCFT2}
\ee
with a constant $C$.

Smearing by the flow equation \eqref{NRflow} does not break the property of the operator as an eigenvector for the mass operator, so the 2-point function of the flowed operator, which we denote by $\phi_m$, further reduces from \eqref{2ptNRflowedOp} to 
\be 
\aver{\phi_m(x_1;\eta_1)\phi_m^\dagger(x_2;\eta_2)} 
= { \exp\[{im \(x_{12}^{-}+ \displaystyle{
(\vec x_{12})^2\over 2(x^+_{12} + 2i\bar m \eta_+) } \)}\]\over \eta_+^{\Delta_\cO}} F_m\({x^+_{12} \over \eta_+};1\) .
\ee
It is important to note the introduction of nonzero $\bar m$ in the flow equation \eqref{NRflow} is needed to resolve the contact singularity in the time ($x^+$) direction.  
Therefore the function $G$ in Section \ref{NRH} reduces to 
\be
G\({2\(x^+ + 2i\bar m \eta_+\)x^{-}+\vec x^2\over \eta_+ }, {x^+ \over \eta_+}\) = \exp\[{im \(x^{-}+ \displaystyle{
\vec x^2\over 2(x^+ + 2i\bar m \eta_+) } \)}\] G_m\({x^+ \over \eta_+}\).
\ee
This suggests that the induced line element \eqref{generalNRgeometry} becomes
\beal{ 
ds^2 =& {\Delta_\cO \over 4\eta^2} d\eta^2  + {-G_m''(0) \over 4\eta^2}(dx^+)^2 +  {-im G_m'(0) \over \eta} dx^+dx^- 
+ m^2 (dx^-)^2 +{-m \over 4\bar m \eta} d\vec x^2.
\label{MassEigenVectorMetric}
}
{The coefficients $G_m'(0) $ and $G_m''(0)$  
are determined from \eqref{2ptFlowConstraintA} and \eqref{2ptFlowConstraintB} as follows.} 
\beal{
{G_m'(0) } =& -{{\Delta_\cO} + {m (d-1) \over 2\bar m} \over 2i(m+\bar m)}, ~~~
{G_m''(0) } = {-(\Delta_\cO+1 + {m (d-1)  \over 2\bar m})G_{m}'(0) - {im (d-1) \over 4\bar m^2} \over 2i(m+\bar m)} . 
} 

{
This geometry becomes a Lifshitz one with $Z=2$ after an appropriate compactification in some parameter region. To realize this, we tune a parameter in the non-relativistic flow as $\bar m = - c m$ with $c > 0$.
Then, by setting 
\begin{equation}
\eta = {-m \over 4\Delta_\cO \bar m}\tau^2 =  {1\over 4c \Delta_\cO }\tau^2,  
\end{equation} 
the above metric becomes 
\beal{ 
ds^2 
=& \Delta_\cO \(c_1{(dx^+)^2 \over \tau^4}  + {d\tau^2 + d\vec x^2 \over \tau^2} \) + m^2 (dx^- +c_2 {dx^+ \over \tau^2 } )^2,
\label{MassEigenVectorMetric2}
}
where $c_1$ and $c_2$ are given by
\beal{
c_1= \Delta_\cO\frac{c^2 (2\Delta_\cO -d+1) +(d-1) (1-c)^2 }{2  (1-c)^2}, ~~~
c_2 = \frac{\Delta_\cO  (2 c \Delta_\cO  -d+1) )}{2 m^2 (1-c)}.
}
Thus,  $c_1$ has to be negative so that the metric \eqref{MassEigenVectorMetric} may be regarded 
as a Lifshitz spacetime with $Z=2$. This is realized in a region specified by   
\be 
(0 <) \ \Delta_\cO < \left({d-1 \over 2}\right)\frac{2c-1}{c^2} , 
\ee
where the upper bound becomes maximum as $ \Delta_\cO < (d-1)/2$ at $c=1$.
}

\section{Holographic geometry from a Lifshitz-type scale-invariant theory}
\label{LifshitzGeometry}

In this section, we consider a Lifshitz-type scale-invariant field theory, which is not necessarily conformally invariant.
There exists a primary scalar operator $O_Z(t,\vec x)$, which transforms under the dilatation as\footnote{The infinitesimal transformation is $\delta_\lambda O_Z(\vec x,t)=-\lambda \left(\Delta_\cO +x_i \partial_i + Z t \partial_t \right) O_Z(\vec x, t)$, where $\Lambda=e^\lambda$.
}
\begin{equation}
O_Z(\Lambda\vec x,\Lambda^Z t)=\Lambda^{-\Delta_\cO} O_Z(\vec x, t)
\label{DilatationLifshitz}
\end{equation}
with a scaling factor $Z$.
Using the invariance  under the translation and the special rotation, the 2-point function of this scalar operator is constrained as 
\begin{equation}
\langle  O_Z(\vec x_1,t_1) O_Z(\vec x_2,t_2) \rangle 
=  { f_Z\left( {{t_{12}}^{2/Z} \over {\vec x_{12}}^2 }\right) \over ({\vec x_{12}}^2)^{\Delta_\cO}}, \quad
\label{2ptLifshitz}
\end{equation}
where $f_Z$ is a function characterized by the original theory.
We assumed it to be smooth in terms of the time and invariant under the time reversal.\footnote{If we do not assume this, then terms such as $d\eta dx^+$ appear in the induced metric.
}

A flow equation in this case should be a diffusion type and compatible with the scaling relation. 
We can easily find out such a flow equation as 
\begin{equation}
\frac{ \partial \phi_Z(\vec x,t;\eta)}{\partial \eta} =( \vec\partial^2 +a Z\eta^{Z-1} \partial_t^2) \phi_Z(\vec x,t ;\eta), \quad
 \phi_Z(\vec x,t; 0) =  O_Z(\vec x,t), 
 \label{eq:flow_L}
\end{equation}
where we here introduced a positive parameter $a$. We call this flow a Lifshitz flow.
The formal solution is given by
\begin{equation}
\phi_Z(\vec x,t;\eta) = e^{\eta\vec\partial^2 +a\eta^Z \partial_t^2} O_Z(\vec x,t) .
\end{equation} 

The 2-point function of the flowed field $\phi_Z$ can be written as
\begin{eqnarray}
 \langle \phi_Z(\vec x_1,t_1; \eta_1) \phi_Z(\vec x_2,t_2; \eta_2)\rangle 
&=&e^{\eta_1\vec\partial^2+ a \eta_1^{Z}\partial_t^2+\eta_2\vec\partial'^2+ a \eta_2^{Z}\partial_t'^2} \langle O_Z(\vec x_1,t_1) O_Z(\vec x_2,t_2) \rangle \nn
&=&e^{\eta_+\vec\partial^2+ a \eta_{Z+} \partial_t^2}\langle O_Z(\vec x_{12},t_{12}) O_Z(\vec 0,0) \rangle
\end{eqnarray}
where we used \eqref{2ptLifshitz} and set $\eta_{Z+}:=\eta_1^Z+\eta_2^Z$. Taking into account \eqref{2ptLifshitz} we can denote this function by $F_Z(\vec x_{12}^2, t_{12}^2; 2^{Z-1}\eta_{Z+}, \eta_+)$. 
Here we used the assumption that the function $f$ in \eqref{2ptLifshitz} is smooth with respect to the time. 

By using the scaling relation \eqref{DilatationLifshitz} the function $F_Z$ satisfies
\begin{equation}
F_Z\left(\Lambda^2x^2, \Lambda^{2Z} t^2; \Lambda^{2Z} 2^{Z-1}\eta_{Z+},\Lambda^2 \eta_+\right)
=\Lambda^{-2\Delta_\cO} F_Z\left(x^2, t^2; 2^{Z-1}\eta_{Z+}, \eta_+\right). 
\end{equation}
Choosing $\Lambda=1/\eta_+^\half$ we find 
\begin{eqnarray}
F_Z(\vec x^2, t^2;\eta_{Z+} ,\eta_+) 
&=&\frac{1}{\eta_+^{\Delta_\cO}} F_Z\left(\frac{\vec x^2}{\eta_+}, \frac{t^2}{\eta_{+}^{Z}}; \frac{2^{Z-1}\eta_{Z+}}{\eta_+^Z},1\right),
\end{eqnarray}
from which the normalized flowed field is given by
\begin{eqnarray}
\sigma_Z(\vec x,t;\eta) &=& \frac{(2\eta)^{\Delta_\cO/2}}{\sqrt{F_Z(0,0; 1,1)}} \phi_Z(\vec x,t;\eta),
\end{eqnarray}
so that
\begin{eqnarray}
\langle \sigma_Z(\vec x_1,t_1; \eta_1) \sigma_Z(\vec x_2,t_2; \eta_2)\rangle 
&=&\left(\frac{2\sqrt{\eta_1\eta_2}}{\eta_+}\right)^{\Delta_\cO} G_Z\left(\frac{\vec x_{12}^2}{\eta_+}, \frac{t_{12}^2}{\eta_+^{Z}}, \frac{2^{Z-1}\eta_{Z+}}{\eta_+^Z}\right),
\end{eqnarray}
where $G_Z(x_1,x_2,x_3) := F_Z(x_1,x_2;x_3,1)/F_Z(0,0;1,1)$.
This function $G_Z$ is constrained by the flow equation:
\be 
\partial_{\eta_1}\langle \phi_Z(\vec x_1,t_1; \eta_1) \phi_Z(\vec x_2,t_2; \eta_2)\rangle 
=( \vec\partial^2 +a Z\eta_1^{Z-1} \partial_t^2)\langle  \phi_Z(\vec x_1,t_1; \eta_1) \phi_Z(\vec x_2,t_2; \eta_2)\rangle 
\ee
which leads to 
\beal{
{-\Delta_\cO } =& {2(d-1)} G_Z^{(1,0,0)}(0,0,1) +a Z {2^{2-Z} } G_Z^{(0,1,0)}(0,0,1), \nn
G_Z^{(0,0,1)}(0,0,1)  
=&  {a  \over 2^{Z-2} } G_Z^{(0,1,0)}(0,0,1) .
}

Non-zero components of the induced metric are calculated as
\beal{ 
g_{\eta\eta}(z) =&{\Delta_\cO \over 4\eta^2 } - {Z(Z-1) \over 4\eta^2}G_Z^{(0,0,1)}(0,0,1), \\
g_{tt}(z) =&{-2 \over (2\eta)^Z} G_Z^{(0,1,0)}(0,0,1), \\
g_{ij}(z) =&{ -\delta_{ij} \over \eta } G_Z^{(1,0,0)}(0,0,1). 
}
Therefore
\beal{ 
ds^2 =& \left({\Delta_\cO \over 4\eta^2} - {Z(Z-1) \over 4\eta^2}G_Z^{(0,0,1)}(0,0,1)\right) d\eta^2 + {-2 \over (2\eta)^Z} G_Z^{(0,1,0)}(0,0,1) (dt)^2  +{-\delta_{ij} G_Z^{(1,0,0)}(0,0,1) \over \eta} dx^idx^j\nn
=& \left(\Delta_\cO - a Z(Z-1) 2^{2-Z}  G_Z^{(0,1,0)}(0,0,1)\right) {d\tau^2 + d\vec x^2\over \tau^2} - {2G_Z^{(0,1,0)}(0,0,1) \over (2\alpha)^Z } { (dt)^2 \over \tau^{2Z}},
}
where we set $\eta := \alpha \tau^2$ with 
\be 
\alpha = {\Delta_\cO +a Z 2^{2-Z} G_Z^{(0,1,0)}(0,0,1)  \over 2(d-1)(\Delta_\cO - a Z(Z-1) 2^{2-Z}  G_Z^{(0,1,0)}(0,0,1)) }.
\ee
This describes a Lifshitz geometry with a general dynamical exponent $Z$.

\section{Conclusion and discussion} 
\label{discussion}
We have extended the construction of holographic geometries by means of the flow equation approach to non-relativistic scale invariant theories. 
After reviewing the construction of the AdS space by using a general CFT both in the Euclidean and the Lorentzian space,
we moved on to the construction of holographic geometries of a non-relativistic CFT and a non-relativistic flow equation. 
As a result we have obtained a hybrid geometry of both Schr\"odinger and Lifshitz geometries as a general holographic space-time in this framework.
Applying this result to specific non-relativistic models, we have reproduced a Schr\"odinger geometry and a Lifshitz one with $Z=2$. 
We have also reproduced a Lifshitz geometry with a general dynamical exponent by smearing an operator of a Lifshitz theory with a suitable modification of the flow equation. 

\medskip 

It would be an interesting problem to realize the seemingly new geometry we called the NR Hybrid geometry in Section \ref{NRH} as a solution of a certain bulk theory. Such a bulk theory may be realized as a usual gravitational theory coupling to matter fields 
in a similar way with Lifshitz and Schr\"odinger geometries (see \cite{Taylor:2015glc} for a review and references therein). 

\medskip 

Although the flow field approach seems to provide new perspective to investigate the holography, there are still gaps to fill in between them. One of them is the relationship between flowed operators in a CFT and bulk operators appearing in the standard AdS/CFT correspondence. 
It may be clear that they are conceptually different, because the 2-point function of a flowed operator does not have contact singularity, while that of a bulk local field has. 
Indeed there is a standard construction of bulk operators from a Lorentzian CFT known as the HKLL construction \cite{Hamilton:2006az}, where bulk operators are obtained by convoluting CFT operators with a certain smearing function. Their striking result is that such a smearing is done over the causally disconnected region to obtain a bulk operator in even dimensional Poincar{\'e} AdS, while smearing is done all over the region for odd dimensional one.  
In Section \ref{LAdS} and appendix \ref{NoncovariantFlow} we smeared a CFT primary operator in Lorentzian flows. In both cases smearing region is basically done all over the space. 
It is important to understand how smearing encodes the causality in the Lorentzian space in the flow equation approach.

\medskip 

In relation to the above, it is also important to investigate the correspondence of excited states between the bulk and boundary in the flow field approach. There are orthodox ways to study bulk geometries corresponding to an excited state (for example \cite{deHaro:2000vlm}), while there is a proposal how to compute a back-reacted geometry by an excited state  in the flow field approach \cite{Aoki:2018dmc}. 
It is intriguing to see whether a resulting induced geometry have desired properties and match one constructed by a different approach. 
  
\medskip 

We hope to come back to these issues in the near future.

\subsection*{Acknowledgments}

We are very grateful to Yoshihiko Abe
for valuable comments and useful discussions.  
The authors thank the Yukawa Institute for Theoretical Physics at Kyoto University. 
{We would like to thank Janos Balog for his valuable comments on the first version of this paper.}
Discussions during the workshop YITP-T-18-04 "New Frontiers in String Theory 2018" 
were useful to complete this work.
S.A. is supported in part by the Grant-in-Aid of the Japanese Ministry of Education, Sciences and Technology, Sports and Culture (MEXT) for Scientific Research (Nos. JP16H03978, JP18H05236),  
by a priority issue (Elucidation of the fundamental laws and evolution of the universe) to be tackled by using Post ``K" Computer, 
and by Joint Institute for Computational Fundamental Science (JICFuS).
The work of K.Y. was supported by the Supporting Program for Interaction-based Initiative 
Team Studies (SPIRITS) from Kyoto University, a JSPS Grant-in-Aid for Scientific Research (B) 
No.\,18H01214. This work is also supported in part by the JSPS Japan-Russia Research Cooperative Program.

\appendix

\section*{Appendix}

\section{Conformal, Schr\"{o}dinger and Lifshitz algebras} 
\label{CSL} 

In this appendix we shall describe how the Schr\"{o}dinger and Lifshitz algebras are embedded 
into the conformal algebra $\mathfrak{so}(2,D)$, and present the transformation laws under 
the Schr\"{o}dinger and Lifshitz symmetries.

\subsection*{Conformal algebra $\mathfrak{so}(2,D)$}
Let us begin with a conformal algebra $\mathfrak{so}(2,D)$ in $D$ dimensional Minkowski spacetime, which is generated by anti-symmetric matrices $M_{AB}$ satisfying
\beal{
[M_{AB},M_{CD}] &=i g_{AC} M_{BD} -i g_{BC} M_{AD} -i g_{AD} M_{BC} +i g_{BD} M_{AC}\,, 
\label{so2D} 
}
Here the indices are the metric components are given by 
\beal{
 A,B,\ldots &= -1, 0, 1, \cdots, D\,, \quad 
 -g_{-1-1}=-g_{00}=g_{11}=\cdots=g_{DD}=1\,. 
\notag 
}
In the following, it is helpful to introduce the light-cone coordinates: 
\beal{x^{\wt\pm} = \frac{1}{\sqrt{2}}(x^{D} \pm x^{-1})\,. 
}
Then the components of $M_{AB}$ can be presented in terms of the conformal basis 
$P_{\mu}$ (translation), $M_{\mu\nu}$ (Lorentz rotation), $D$ (dilatation) 
and $K_{\mu}$ (special conformal) as follows:  
\begin{equation}
(M_{AB})=
\bordermatrix{
  &  \nu & \wt+ & \wt- \cr
\mu & M_{\mu\nu} & P_{\mu} & K_{\mu} \cr
\wt+ & -P_{\nu}& 0 & -D \cr
\wt- & -K_{\nu} & D & 0 
}. 
\end{equation}
Now the commutation relation in \eqref{so2D} can be decomposed into the following standard form: 
\beal{
[M_{\mu\nu},M_{\rho\delta}]=&i g_{\mu\rho} M_{\nu\delta} -i g_{\nu\rho} M_{\mu\delta} -i g_{\mu\delta} M_{\nu\rho} +i g_{\nu\delta} M_{\mu\rho},
\\
\left[M_{{\mu}{\nu}},P_{{\rho}}\right] =& i g_{{\mu}{\rho}} P_{{\nu}} - i g_{{\nu}{\rho}} P_{{\mu}}, \quad 
\left[M_{{\mu}{\nu}},K_{{\rho}}\right] =i g_{{\mu}{\rho}} K_{{\nu}} - i g_{{\nu}{\rho}} K_{{\mu}},\\
\left[K_{{\mu}}, P_{{\nu}}\right] =&i g_{{\mu}{\nu}} D + i M_{{\mu}{\nu}}, \\
\left[D, M_{{\mu}{\nu}} \right] =& 0, \quad 
\left[D,P_{{\nu}}\right] = iP_{\nu}, \quad 
\left[D,K_{{\nu}}\right] =-iK_{\nu}.
}

\subsection*{Schr\"odinger algebra from $\mathfrak{so}(2,D)$}

To see the Schr\"odinger algebra as a subalgebra of $\mathfrak{so}(2,D)$\,, 
it is useful to introduce another couple of the light-cone coordinates $x^{\pm}$ 
with $x^{0}$ and $x^{D-1}$~: 
\beal{
 x^{\pm} = \frac{1}{\sqrt{2}} (x^{D-1} \pm x^{0})\,.
 } 
Then the generators of $\mathfrak{so}(2,D)$ can be displayed as  
 \begin{equation}
(M_{AB})=
\bordermatrix{
 & + & -  & j & \wt+ & \wt- \cr
+ & 0 & M_{+-}&- M_{j+} & P_{+}& K_{+}\cr
- & M_{-+} & 0&- G_j & P_{-}& K_{-}\cr
{i}&M_{i+} &G_i  & M_{ij} & P_i & K_i \cr
\wt+ & -P_{+}& -P_{-}& -P_j& 0 & -D \cr
\wt- & -K_{+}& -K_{-}& -K_j & D & 0 
}\,.
\end{equation}
It is significant to notice that one can find out a subalgebra by dropping 
the generators $M_{i+}$\,, $K_i$ and $K_{+}$\,.  
By introducing the following notation 
\be
H \equiv P_{+}, \quad   
M \equiv P_{-}, \quad 
K \equiv K_{-}, \quad 
\cD =D + M_{-+}\,, 
\ee
the subalgebra is given by  
\beal{
[M_{ij},M_{kl}]=& i g_{ik} M_{jl} -i g_{jk} M_{il} -i g_{il} M_{jk} +i g_{jl} M_{ik},  \notag \\
\left[M_{ij},P_{{k}}\right] =& i g_{i{k}}P_{j} - i g_{j{k}}P_{i}, \quad 
\left[M_{ij},G_{{k}}\right] = i g_{i{k}}G_{j} - i g_{j{k}}G_{i} \notag \\
\left[G_{i},P_{j}\right] =& ig_{ij} M, \quad 
\left[H, G_{i} \right] =i P_i, \quad 
\left[H,K\right] = -i \cD,  \notag \\
\left[\cD,P_{j}\right] =& iP_j, \quad 
\left[\cD,G_{j}\right] = - iG_j, \quad 
\left[\cD,H\right] = 2iH, \quad 
\left[\cD,K\right] = -2iK, \quad \label{A11}
}
and the other commutation relations vanish. This is nothing but the Schr\"odinger algebra\footnote{This is the Schr\"odinger algebra with the dynamical critical exponent $z_c=2$\,. One may consider an arbitrary value of $z_c$\,. But except $z_c \neq2$ (and 1), the special conformal generator $K$ must be excluded so as to close the algebra. }. 
This algebra is composed of $H$ (time translation), $P_i$ (spatial translation), 
$M_{ij}$ (spatial rotation), $G_i$ (Galilean boost), $\mathcal{D}$ (anisotropic dilatation)
$K$ (special conformal) and $M$ (mass operator).  

\medskip 

As one can notice from the commutation relations in \eqref{A11}, the scale transformation associated 
with $\mathcal{D}$ is {\it anisotropic} like 
\begin{eqnarray}
t \rightarrow \Lambda^2 t\,, \quad x^i \rightarrow \Lambda x^i \qquad (\Lambda:~\mbox{a real constant parameter})\,. 
\end{eqnarray}
This is a characteristic of the Schr\"odinger algebra.

\subsection*{Lifshitz algebra from Schr\"odinger algebra}
The Lifshitz algebra is embedded as a subalgebra of the Schr\"odinger algebra when the dynamical exponent is two. 
This embedding can be seen by dropping off the generators $G_i$, $K$ and $M$ from the Schr\"odinger algebra (\ref{A11}). The resulting commutation relations are given by 
\beal{
[M_{ij},M_{kl}]=& i g_{ik} M_{jl} -i g_{jk} M_{il} -i g_{il} M_{jk} +i g_{jl} M_{ik},  \notag \\
\left[M_{ij},P_{{k}}\right] =& i g_{i{k}}P_{j} - i g_{j{k}}P_{i}, \notag \\
\left[\cD,P_{j}\right] =& iP_j, \quad \left[\cD,H\right] = 2iH\,.  \label{A12}
}
In total, this algebra is composed of $H$ (time translation), $P_i$ (spatial translation), 
$M_{ij}$ (spatial rotation) and $\mathcal{D}$ (anisotropic dilatation).
To recover the algebra with an arbitrary value of the dynamical exponent $Z$, 
the commutation relation involving the dilatation and Hamiltonian 
should be modified as 
\be
\left[\cD,H\right] = ZiH\,.  \label{A13}
\ee

\section{Lorentz non-invariant flow equation} 
\label{NoncovariantFlow}

In this appendix we present a different method to flow a primary operator in Lorentzian CFT. Although this approach breaks the manifest Lorentz invariance, it has a virtue to obtain a well-defined flowed operator in the Lorentzian space. 

The method is to introduce another flow parameter $\eta_t$ specially for the time direction as follows.   
\be 
{\partial \over \partial \eta_t}\phi_0 = -\partial_t^2 \phi_0, \quad 
{\partial \over \partial \eta}\phi_0 = \sum_i \partial_i^2 \phi_0, \quad 
\ee
If the flow parameters are in the region that 
\be 
\eta_t < 0 , \quad \eta > 0, 
\ee
then the flow equation has a well-defined solution 
\beal{  
\phi_0(x;\eta_t, \eta) 
=& e^{-\eta_t \partial_t^2 + \eta \partial_i^2} O_0(x)
=\int d^Dx' K_0(x,x';\eta_t, \eta) O_0(x') 
}
where 
\beal{
K_0(x,x';\eta_t, \eta)
=&   {e^{ {(t-t')^2 \over 4\eta_t} } \over (4\pi\eta_t)^\half} {e^{ - {(x^i-x'^i)^2 \over 4\eta} } \over (4\pi\eta)^{D-1\over 2}}. 
}
Then the two point correlation function of the flowed operator is written as
\beal{ 
\aver{\phi_0(x_1;\eta_{t1}, \eta_1)\phi_0(x_2;\eta_{t2}, \eta_2)} 
=& e^{-\eta_t \partial_t^2 + \eta \partial_i^2-\eta'_t \partial_t'^2 + \eta' \partial_i'^2} f_0(x_{12}^2) \nn
=& e^{-\eta_{t+} \partial_t^2 + \eta_+ \partial_i^2} f_0(x_{12}^2), 
\label{F_02}
}
where we used the notation in the main text. 
This is a function of $t_{12}^2,\vec x_{12}^2, \eta_{t+},  \eta_{+}$, which we denote by $F_0(t_{12}^2,\vec x_{12}^2;\eta_{t+},  \eta_+)$. 
By using the scaling relation we find  
\be 
F_0(t^2, \vec x^2;\eta_t,\eta) = {1\over \eta^{\Delta_\cO}} F_0\left({t^2 \over \eta},{\vec x^2 \over \eta};{\eta_t \over \eta},1\right).
\ee 
In particular 
\beal{
\aver{\phi_0(x;\eta_t, \eta)^2} 
=&F_0(0,0;2\eta_t, 2\eta)
= {1\over (2\eta)^{\Delta_\cO}} F_0(0,0;{\eta_t\over \eta},1)
}
Furthermore we have
\begin{eqnarray}
F_0^{(1,0,0,0)}\left({t^2 \over \eta},{\vec x^2 \over \eta};{\eta_t \over \eta},1\right)
&=& (-1){ \eta^{\Delta_\cO}\over \eta} e^{-\eta_{t} \partial_t^2 + \eta \partial_i^2} f_0^\prime(x_{12}^2)
= - F_0^{(0,1,0,0)}\left({t^2 \over \eta},{\vec x^2 \over \eta};{\eta_t \over \eta},1\right) .
\label{dF_0}
\end{eqnarray}

Therefore the two point correlation function of the normalized field is given by
\beal{ 
\aver{\sigma_0(x_1;\eta_{t1}, \eta_1) \sigma_0(x_2;\eta_{t2}, \eta_2) } 
=\left({\sqrt{4\eta_1\eta_2}\over \eta_+}\right)^{\Delta_\cO} { F_0({t_{12}^2 \over \eta_+},{\vec x_{12}^2 \over \eta_+};{\eta_{t+} \over \eta_+},1) \over \sqrt{F_0(0,0;{\eta_{t1}\over \eta_1},1) F_0(0,0;{\eta_{t2}\over \eta_2},1)} }.
}
The right hand side is convergent and smooth for $\eta_{t1}, \eta_{t2} < 0, \eta_1, \eta_2>0$. 
Therefore we perform the analytic continuation in terms of $\eta_{t1}, \eta_{t2}$ from the negative region to the positive value with $\eta_t\to \eta, \eta_{t2}\to\eta_2$. Then the right-hand side yields a smooth function of ${t_{12}^2 \over \eta_+}$ and ${\vec x_{12}^2 \over \eta_+}$, which we denote as
$\({\sqrt{4\eta_1\eta_2}\over \eta_+}\)^{\Delta_\cO} G_0\({t_{12}^2 \over \eta_+},{\vec x_{12}^2 \over \eta_+}\right) $ with $G_0(\vec0) =1$.
We thus obtain 
\beal{ 
\aver{\sigma_0(x_1;\eta_1) \sigma_0(x_2;\eta_2) } 
=& \left({\sqrt{4\eta_1\eta_2}\over \eta_+}\right)^{\Delta_\cO} G_0\left({t_{12}^2 \over \eta_+},{\vec x_{12}^2 \over \eta_+}\right)
}
where we set  
\be 
\aver{\sigma_0(x_1;\eta_1) \sigma_0(x_2;\eta_2) } 
:= \lim_{\eta_{t1}\to \eta_1 \atop \eta_{t2}\to\eta_2} \aver{\sigma_0(x_1;\eta_{t1}, \eta_1) \sigma_0(x_2;\eta_{t2}, \eta_2) }. 
\ee

By using this the induced metric is computed as 
\beal{ 
g_{\eta\eta}(z) =& {\Delta_\cO \over 4\eta^2 }, ~~~
g_{ij}(z) ={-\delta_{ij}\over \eta}G_0^{(0,1)}(\vec0), ~~~ g_{tt}(z) = {-1\over \eta}G_0^{(1,0)}(\vec0).
}

From the flow equation we find 
\beal{
{-{\Delta_\cO} } F_0(0,0;1,1)-\frac{1}{\eta}F_0^{(0,0,1,0)}(0,0;1,1)
=& 2(D-1)  F_0^{(0,1,0,0)}(0,0;1,1), \nn
\frac{1}{\eta}F_0^{(0,0,1,0)}(0,0;1,1) =& -2F_0^{(1,0)}(0,0;1,1), 
} 
which leads to  
\beal{
{-{\Delta_\cO} } = 2(D-1) G_0^{(0,1)}(\vec0) -2 G_0^{(1,0)}(\vec0).
} 
By using \eqref{dF_0} we find $G_0^{(0,1)}(\vec0) = - G_0^{(1,0)}(\vec0)$, which gives $G_0^{(0,1)}(\vec0) = {-{\Delta_\cO} \over 2D}.$
Using these relations we obtain the induced line element as 
\begin{eqnarray}
\dd s^2 &=& \Delta_\cO \frac{-\dd t^2 + \dd\vec x^2 + \dd\tau^2}{\tau^2} ,
\label{eq:AdS_L}
\end{eqnarray}
where we set $\tau^2 = 2D\eta.$ This is the Lorentzian AdS metric.

\section{Transformation of the flowed field}
\label{formula}

\subsection{Useful formulas}
 We here collect some useful formulas to calculate transformation properties of the flowed field.
 
 \begin{eqnarray*}
 e^{\eta \vec \partial^2} x^i &=& \left(x^i + 2\eta\partial^i\right)  e^{\eta \vec \partial^2}, \\
 e^{\eta \vec \partial^2} x^i x^j &=& \left( x^i + 2\eta\partial^i\right) \left( x^j + 2\eta\partial^j\right) e^{\eta \vec \partial^2}= \left( x^i x^j + 2\eta(x^i\partial^j+x^j\partial^i +\delta_{ij}) +4\eta^2 \partial^i\partial^j\right) e^{\eta \vec \partial^2} ,\nonumber \\
 e^{\eta \vec \partial^2} \vec x^2 &=& \left( \vec x^2 + 4\eta \vec x\cdot \vec \partial + 2(d-1)\eta +4\eta^2 \vec \partial^2\right) e^{\eta \vec \partial^2} ,\nonumber 
 \end{eqnarray*}
 \begin{eqnarray*}
 e^{2\eta \partial_+\partial_-} x^+ &=& \left(x^+ + 2\eta\partial_-\right) e^{2\eta \partial_+\partial_-}  , \\
 e^{2\eta \partial_+\partial_-}  (x^+)^2  &=& \left( x^+ + 2\eta\partial_-\right)^2 e^{2\eta \partial_+\partial_-} = \left( (x^+)^2+ 4\eta x^+ \partial_- +4\eta^2 \partial_-^2\right) e^{2\eta \partial_+\partial_-} ,\nonumber \\
 e^{2\eta \partial_+\partial_-} x^- &=& \left(x^- + 2\eta\partial_+\right) e^{2\eta \partial_+\partial_-}  , \\
 e^{2\eta \partial_+\partial_-}  (x^-)^2  &=& \left( x^- + 2\eta\partial_+\right)^2 e^{2\eta \partial_+\partial_-} = \left( (x^-)^2+ 4\eta x^- \partial_+  +4\eta^2 \partial_+^2\right) e^{2\eta \partial_+\partial_-} ,\nonumber \\
 e^{2\eta \partial_+\partial_-} x^+ x^- &=&  \left(x^+ + 2\eta\partial_-\right)  \left(x^- + 2\eta\partial_+\right) e^{2\eta \partial_+\partial_-}  \nonumber \\
 &=&\left(x^+x^- +2\eta(x^+\partial_+ +x^-\partial_- +1) + 4\eta^2 \partial_+\partial_-\right) e^{2\eta \partial_+\partial_-} , 
 \end{eqnarray*}
 \begin{eqnarray*}
 e^{\epsilon \partial_-^2} x^- &=& \left(x^- + 2\epsilon\partial_-\right) e^{\epsilon \partial_-^2} , \\
 e^{\epsilon \partial_-^2} (x^-)^2 &=& \left(x^- + 2\epsilon\partial_-\right)^2 e^{\epsilon \partial_-^2} 
 \left((x^-)^2 + 2\epsilon(2x^-\partial_- +1) +4\epsilon^2 \partial_-^2\right) e^{\epsilon \partial_-^2} ,
 \\
 e^{\epsilon \partial_-^2} x^+x^- &=& x^+ \left(x^- + 2\epsilon\partial_-\right) e^{\epsilon \partial_-^2} ,
 \end{eqnarray*}
 \begin{eqnarray*}
 e^{2\eta \partial_+\partial_- +\epsilon\partial_-^2}  x^- &=& \left(x^- + 2\eta\partial_+\ + 2\epsilon\partial_-\right) 
  e^{2\eta \partial_+\partial_- +\epsilon\partial_-^2}  , \\
 e^{2\eta \partial_+\partial_- +\epsilon\partial_-^2}   (x^-)^2  &=& \left( (x^-)^2 + 4\eta x^-\partial_+ + 4\eta^2 \partial_+^2 +4\eta\epsilon\partial_+\partial_- +2\epsilon(2x^-\partial_-+1) + 4\epsilon^2\partial_-^2\right)
 e^{2\eta \partial_+\partial_- +\epsilon\partial_-^2}, \\
 e^{2\eta \partial_+\partial_- +\epsilon\partial_-^2}  x^+ x^- &=&  \left(x^+ x^- + 2\eta(x^- \partial_- +x^+\partial_+ +1) +4\eta^2\partial_+\partial_- +4\eta\epsilon \partial_-^2 + 2\epsilon x^+\partial_-\right)
 e^{2\eta \partial_+\partial_- +\epsilon\partial_-^2} ,
 \end{eqnarray*}
  \begin{eqnarray*}
 e^{a\partial_+ } (x^+)^n &=& \left(x^+ +a\right)^n e^{a \partial_+ } .
 \end{eqnarray*}
 
\subsection{General transformation properties}
Under the conformal transformation, the general flowed field $\phi_{\epsilon,m}(x;\eta)$ transforms as
\begin{eqnarray}
{\delta^\prime}^{\rm conf}\phi_{\epsilon,m}(x;\eta) &:=& e^{\eta(\vec \partial^2 + 2\partial_+\partial_-)+\epsilon\partial_-^2 + i \eta 2\bar m \partial_+}\, \delta^{\rm conf}O(x) = \delta^{\rm conf}\phi_{\epsilon,m}(x;\eta) +\Delta^{\rm conf} \phi_{\epsilon,m}(x;\eta) , ~~~~~~\\
\Delta^{\rm conf} \phi_{\epsilon,m}(x;\eta)&:=& \left(\delta^\epsilon + \delta^\eta+\delta^{\eta^2}+\delta^{\epsilon\eta}\right) \phi_{\epsilon,m}(x;\eta), 
\end{eqnarray}
where 
\begin{eqnarray}
\delta^{\rm conf} &:=& -\left[ a\cdot \partial + \omega^{\mu}{}_\nu \bar x^\nu \partial_\mu + \lambda (\bar x\cdot\partial + \Delta_\cO) + \bar x^2 b\cdot\partial - 2 b\cdot \bar x (\bar x\cdot \partial + \Delta_\cO) \right], \\
\delta^\eta &:=& 2\eta\left( 2 b\cdot \bar x -\lambda\right) ( \partial_\eta -2i\bar m \partial_+) 
-2\eta (d-1-2\Delta_\cO) b\cdot \partial,  \\
\delta^\epsilon &:=& 2\epsilon \left[ (2 b\cdot \bar x -\lambda-\omega^{-+}) \partial_-^2 +2(\bar  x\cdot\partial +\Delta_\cO+1) b^+\partial_- -\omega^{i+}\partial_i\partial_-  -2 \bar x^+\partial_- b\cdot \partial \right],~~~~~ \\
\delta^{\eta^2}&:=& 4\eta^2  b\cdot \partial ( \partial_\eta -2i\bar m \partial_+) , \\
\delta^{\epsilon\eta} &:=& 8 \epsilon\eta b^+\partial_-( \partial_\eta -2i\bar m \partial_+) , \\
\delta^{\epsilon^2} &:=& 8\epsilon^2 b^+\partial_-^3,
\end{eqnarray}
and $\bar x  = (\vec x, x^++2i\bar m\eta, x^-)$
with $a\cdot\partial := a^i\partial_i + a^+\partial_+ + a^-\partial_-$.

\end{document}